\documentclass[
aps,%
12pt,%
final,%
notitlepage,%
oneside,%
onecolumn,%
nobibnotes,%
nofootinbib,%
superscriptaddress,%
noshowpacs,%
centertags]%
{revtex4}

\begin{document}

\pagenumbering{arabic}

\title{
Rovibronic energy levels for triplet electronic states
of molecular deuterium}

\author{B.~P.~Lavrov}
\email{lavrov@pobox.spbu.ru}
\author{I.~S.~Umrikhin}
\affiliation{
Faculty of Physics, St.-Petersburg State University, \\
St.-Petersburg, 198904, Russia}

\begin{abstract}

An optimal set of 1050 rovibronic energy levels for 35 triplet 
electronic states of $D_2$ has been
obtained by means of a statistical analysis of all available 
wavenumbers of triplet-triplet rovibronic transitions studied in emission,
absorption, laser and anticrossing spectroscopic experiments of
various authors. We used a new method of the analysis (Lavrov,
Ryazanov, JETP Letters, 2005), which does not need any \it
a~priori \rm assumptions concerning the molecular structure being
based on only two fundamental principles: maximum likelihood and
Rydberg-Ritz. The method provides the opportunity to obtain the
estimation of uncertainties of experimental wavenumbers independent
from those presented in the original papers. 234 from 3822
published wavenumber values were found to be spurious, while the 
remaining set of the data may be divided into 19 subsets of uniformly 
precise data having close to normal distributions of random errors 
within the subsets.
New wavenumber values of 125 questionable 
lines were measured in the present work (20-th subset).
Optimal values of the rovibronic levels were obtained from the experimental 
data set consisting of 3713 wavenumber values (3588 old and 125 new).  
The unknown shift between levels of ortho- and para-
deuterium was found by least squares analysis of the
$a^3\Sigma_g^+$, $v = 0$, $N = 0 \div 18$ rovibronic levels with 
odd and even
values of N. All the energy levels were obtained relative to the
lowest vibro-rotational level ($v = 0$, $N=0$) of 
the $a^3\Sigma_g^+$ electronic state  
and presented in tabular form together with standard deviations (SD) 
of the semi-empirical determination. New energy level values 
differ significantly from those available in literature.
\end{abstract}

\maketitle

\section*{Introduction}

It is well known that diatomic hydrogen, being the simplest neutral
molecule, has a most sophisticated emission spectrum. The hydrogen band
spectrum, caused by spontaneous emission due to
electronic-vibro-rotational (rovibronic) transitions, 
does not show a visible, easily recognizable band structure, but
has the appearance of a multiline atomic spectra. The peculiarity of
molecular hydrogen and its isotopic species - abnormally small
nuclear masses - leads to high values of vibrational and
rotational constants and large separation between vibrational and
rotational levels of various excited electronic states. As a
result, various rovibronic spectral lines belonging to different
branches, bands and band systems are located in the same 
spectral regions, leading to the overlap of various band
systems, bands and branches, as well as the mixing of rovibronic spectral
lines having different origins. The small nuclear masses 
stimulate a breakdown of the Born-Oppenheimer approximation due
to electronic-vibrational and electronic-rotational perturbations
having both regular and irregular character; this combination seriously
complicates the interpretation of the spectra of hydrogen isotopomers
and the unambiguous identification of rovibronic spectral lines.
Symmetry rules for permutation of identical nuclei in homonuclear
isotopomers (H$_2$,D$_2$ and T$_2$) cause the known effect of the 
intensity alternation of neighbouring lines within the rotational 
structure of bands
due to the alternation in degeneracy of successive rotational
levels with odd and even values of rotational quantum number
(e.g. 1:2 in the case of D$_2$). This effect also masks the
visible structure of branches resulting in serious
additional difficulties for identification of rovibronic 
spectral lines. Thus, most of the lines in the optical spectra of hydrogen
isotopomers have not yet been assigned in spite of tremendous
efforts by spectroscopists over the previous century \cite{Richardson,
Dieke01, Dieke02, Crosswhite}. As an example, in the latest
compilation  of experimental data for molecular deuterium D$_2$ 
\cite{Crosswhite}, the working list of 27488 measured lines
contains only 8243 assignments. These assignments were obtained by
traditional methods of analysis using wavenumber combination
differences (method of common differences) 
and Dunham series expansions \cite{Richardson,Dieke02}, sometimes 
together with comparison of molecular constants obtained for
different isotopic species \cite{Dieke01}. Later the results of
\it ab~initio \rm and semi-empirical calculations were taken into
account \cite{Crosswhite}.

Currently, available information concerning triplet rovibronic energy 
level values of
$D_2$ molecule exists in the form of the list of molecular
constants for Dunham expansions in the NIST database \cite{Nist},
and tables of rovibronic levels obtained in Ref.
\cite{Crosswhite}.

In the case of the hydrogen molecule 
the Dunham coefficients are known to provide a very poor description
of the rovibronic energy level values \cite{LPU}. The data reported in Ref.
\cite{Crosswhite}, in general, give a rather good description of the
$D_2$ spectrum (about $0.05$ cm$^{-1}$), but they are also not free 
from criticism. The
method of the analysis used in Ref. \cite{Crosswhite} is based on
the common use of the combination differences, some selected
wavenumbers for certain transitions and one by one multistage treating of
separate branches, bands and band systems. 
The sequence of the steps chosen in Ref.
\cite{Crosswhite} is not the only possible analytical arrangement.
Therefore, the data thus
obtained can not be considered as an optimal set of levels providing
the best description of observed wavenumber values.
It should also be mentioned that after publication of Ref.
\cite{Crosswhite}, new experimental data on the wavenumbers
appeared \cite{DabrHerz, Davies}.

Recently, a new method of statistical analysis of the rovibronic
transition wavenumbers has been proposed \cite{LavrovRiazanovJetf}
and successfully applied for the derivation of rovibronic level values
of the singlet states of the $BH$ molecule \cite{BHterms}
and some triplet states of $H_2$ \cite{ALMU2008}.
The method is based on only two
fundamental principles: Rydberg-Ritz and maximum likelihood. 
This approach
differs from known techniques in several aspects: 
1) does not need any assumptions concerning an internal
structure of a molecule; 
2) no intermediate parameters, such as molecular constants used in 
the traditional approach, are used; 
3)a one-stage optimization procedure can be used for all available 
experimental data obtained for various band systems, by various methods 
and authors, and in various works; 
4) provides the opportunity of a rational selection of the experimental
data in an interactive mode (thus allowing the user the option to eliminate
obvious errors, to revise incorrect line identifications, and to
compare various sets of experimental data for mutual consistency);
5) gives an opportunity of independent estimation of experimental 
errors by means of the analysis of the shape of error distributions; 
6) provides an optimal set of rovibronic level values as well as the
uncertainties of their determination (standard deviations SD) 
caused only by the quantity and quality of existing
experimental data \cite{LavrovRiazanovOiS}.

The goal of the present paper is to report the results of applying
the new method \cite{LavrovRiazanovJetf} for statistical analysis
of the rovibronic spectral line wavenumbers of triplet band systems
and the determination of the optimal set of rovibronic energy levels 
for
all known 35 triplet electronic states of molecular deuterium:
$a^3\Sigma_g^+$, $c^3\Pi_u^+$, $c^3\Pi_u^-$, $d^3\Pi_u^+$,
$d^3\Pi_u^-$, $e^3\Sigma_u^+$, $f^3\Sigma_u^+$, $g^3\Sigma_g^+$,
$h^3\Sigma_u^+$, $i^3\Pi_g^+$, $i^3\Pi_g^-$, $j^3\Delta_g^+$,
$j^3\Delta_g^-$, $k^3\Pi_u^+$, $k^3\Pi_u^-$, $n^3\Pi_u^+$,
$n^3\Pi_u^-$, $p^3\Sigma_g^+$, $q^3\Sigma_g^+$, $r^3\Pi_g^+$,
$r^3\Pi_g^-$, $s^3\Delta_g^+$, $s^3\Delta_g^-$, $u^3\Pi_u^+$,
$u^3\Pi_u^-$, $(7p)^3\Pi_u^+$, $(7p)^3\Pi_u^-$, $(8p)^3\Pi_u^+$,
$(8p)^3\Pi_u^-$, $(9p)^3\Pi_u^+$, $(9p)^3\Pi_u^-$,
$(6d)^3\Sigma_g^+$, $(7d)^3\Sigma_g^+$, $(8d)^3\Sigma_g^+$ and
$(9d)^3\Sigma_g^+$.

\section*{The method of analysis and determination of optimal set of rovibronic energy levels}

The method of assessment is based on the minimization of the weighted
mean-square deviation between observed $\nu^{n''v''N''}_{n'v'N'}$
and calculated (as differences of adjustable energy levels
$E_{n''v''N''}, \, E_{n'v'N'}$) values of rovibronic line
wavenumbers, or the sum

\begin{equation}
r^2=\sum_{\nu^{n''v''N''}_{n'v'N'}}
 \left[\frac{(E_{n''v''N''}-E_{n'v'N'})-\nu^{n''v''N''}_{n'v'N'}}
            {\sigma_{\nu^{n''v''N''}_{n'v'N'}}}\right]^2,\label{neviazka}
\end{equation}

where $n$ denotes the electronic state, $v$ is the vibrational quantum number,
and $N$ is the rotational quantum number of total angular momentum of a molecule
excluding spins of electrons and nuclei. Upper and lower rovibronic 
levels are marked by double and single primes, respectively.   
 The values $\sigma_{\nu^{n''v''N''}_{n'v'N'}}$ are experimental 
root-mean-square (RMS) errors (one SD), and the summation is over 
all inputed experimental data.
The $\nu^{n''v''N''}_{n'v'N'}$ and
$\sigma_{\nu^{n''v''N''}_{n'v'N'}}$ values are the input
data, while the optimal set of energy levels $E_{nvN}$ and the
calculated standard deviations  of their determination 
$\sigma_{E_{nvN}}$ are
output data from the computer code. Due to the linearity of the
equations used, the optimization problem reduces to the solution 
of a system
of linear algebraic equations \cite{LavrovRiazanovJetf}.

It should be stressed, that for a realization of the method
it is necessary to know the proper estimation of the SD for every 
experimental datum, or be able to divide the whole data set 
into a finite number of groups (subsets, samples) of data measured 
with the same precision.
Then all data belonging to each subset may be characterized by
a certain common value of SD. In the present study we assumed that
systematic errors are eliminated (except occasional spurious data), and
random errors should possess the normal (Gaussian) distribution function. 

The method requires the known RMS errors of the wavenumbers 
in the process of determination the optimal level values  for three
purposes. Firstly, weightening of the input data depends on these
values. Secondly, the rational selection of input data is based on
a comparison of an experimental RMS uncertainty (common for a sample of data 
under the study) with differences between experimental wavenumbers of the lines 
(included into this sample) and those
calculated from the optimal values of the energy levels . 
Finally, estimates of uncertainties of the optimal level values, 
$\sigma_{E_{nvN}}$ are calculated by taking into account the input 
wavenumber RMS errors $\sigma_{\nu}$.

Unfortunately only a few original works with experimental wavenumber
data contain information on the accuracy of the listed values.
And even in these cases the error estimations are of uncertain reliability.
Therefore, it was necessary to obtain independent estimates of 
the wavenumber errors $\sigma_{\nu}$. Our method provides the opportunity 
to obtain the estimates of uncertainties of experimental wavenumbers 
independent from those presented in the original papers.

\section*{Statistical analysis}

The method of analysis consists of analyzing the distribution
function $F(\xi)$ of the variable

\begin{equation}
\xi^{n''v''N''}_{n'v'N'} =
  \frac{\nu^{n''v''N''}_{n'v'N'} -
        (\tilde E_{n''v''N''}-\tilde E_{n'v'N'})}
       {\sigma_{\nu^{n''v''N''}_{n'v'N'}}}.\label{xi}
\end{equation}

Energy level values $\tilde E_{n''v''N''}$ and $\tilde E_{n'v'N'}$
are obtained by the minimization of Eq.(\ref{neviazka}) without using
the $\nu^{n''v''N''}_{n'v'N'}$ wavenumber value. In favorable
cases this distribution should be close to the normal distribution
with an average value equal to zero and variance equal to unity, namely

\begin{equation}
F_{0}(\xi) = \frac{1}{\sqrt{2 \pi}}
  \int \limits_{-\infty}^{\xi} e^{-\frac{x^2}{2}} dx.\label{norm_f}
\end{equation}

The analysis of the shape of the distribution function $F(\xi)$
for various groups (samples) of experimental data provides
an opportunity to estimate the RMS errors of experimental data reported
in various works by the rational selection of experimental data
and determination of proper values of experimental uncertainties
$\sigma_{\nu^{n''v''N''}_{n'v'N'}}$. Experimental data which
have shown unreasonably large deviations ($\xi$ larger then 3)
were excluded as spurious results. Then the value of 
an experimental RMS uncertainty (unique for all lines belonging to 
the same sample) was adjusted in such 
a way that the distribution function $F(\xi)$ of a certain sample
becomes close to the normal distribution function 
represented by Eq.(\ref{norm_f}).

All available values of rovibronic transition wavenumbers studied
in emission, absorption, laser and anticrossing spectroscopic
experiments of various authors were analyzed \cite{DiekeBlue, Dieke, 
GloersenDieke, DiekePorto, Crosswhite,
FreundMiller, DabrHerz, Davies} . Fine and hyperfine
structures of rotational levels were ignored. In the case of
resolved and partly resolved fine structure, the wavenumber
of the strongest component was used.

The statistical analysis of the 3822 published wavenumber
values has shown that 234 experimental data (for 228 spectral 
lines) should be excluded from further consideration as being spurious.
The list of spurious data is presented in Table \ref{dellines}.
The remaining set of 3588 wavenumber values 
could be divided into 19 subsets (samples) of uniformly 
precise data having close to the normal distribution functions for
random experimental errors. All the wavenumber data within each sample
may be characterized by the same value of the RMS uncertainty (one SD).
The uncertainty estimates obtained in the present paper for various bands
Are shown in Table \ref{tabband}.

\section*{Experimental determination of wavenumbers for questionable lines}

135 of the 228 questionable lines are located within the wavelength 
range available for us (430-730 nm). Therefore we decided
to provide an independent experimental determination of their wavenumbers.
For that purpose we used the emission spectra of $D_2$ obtained 
during our previous studies of translational and rotational temperatures 
in hydrogen and deuterium containing plasmas \cite{LKOR1997}.  
A detailed description of the experimental setup has been reported 
elsewhere \cite{AKKKLOR1996}. Capillary arc discharge lamps DDS-30 
described in Ref. \cite{LT1982} have been used as light sources. They were 
filled with about 6 Torr of spectroscopically pure $D_2$ + $H_2$ (9:l)
mixture. The range of the discharge current was from
50 to 300 mA (current densities $j=1.6 \div 10$ $A/cm^2$). 
Light from the axis of the plasma inside the capillary
was directly focused by an achromatic lens on the entrance slit of the
Czerny-Turner type l m double monochromator (Jobin Yvon, U1000).
The intensity distribution in the focal plane of the spectrometer was
recorded by a cooled CCD matrix detector of the Optical Multichannel
Analyser (Princeton Appl. Res., OMA-Vision-CCD System).

Assignments and wavelength values from Ref. \cite{Crosswhite} were 
used for identification of $D_2$ spectral lines in the spectrum. 
These values show a certain spread around monotonic dependence of 
the wavenumber from the distance along the direction of dispersion 
in the focal plane of the spectrograph. For the majority of strong, 
unblended lines, this scatter was within the error
bars of $0.05$ cm$^{-1}$ reported in Ref. \cite{Crosswhite}.

Assuming the dispersion of the spectrograph to be a monotonic
function of the coordinate in the focal plane of the spectrograph,
new wavenumber values for 125 of the questionable spectral lines were 
obtained by polynomial approximation of the dispersion curve.
The new data thus obtained are listed in Table \ref{NewLines}
together with the data reported earlier. One may see that new
experimental wavenumber values are significantly different from 
those from the literature. Moreover, our data are in much better
agreement with Rydberg-Ritz combination principle.
The intensities of the remaining 10 questionable lines were too weak 
to be detectable in the experimental conditions.

Statistical analysis of the deviations given by Eq.(\ref{xi}) have shown 
that the 125 new wavenumber values represent the sample of the
experimental data with the distribution function $F(\xi)$ close to 
normal distribution, corresponding to the value of standard deviation
$\sigma_{\nu}=0.06$ cm$^{-1}$ common for all 125 lines. Such a relatively 
high value of the experimental uncertainty
arises because most of questionable lines are
quite weak and are partly blended with stronger lines. Nevertheless, our new 
data are in better agreement with the optimal set of energy levels, 
obtained by the minimization of Eq.(\ref{neviazka}) with the input data set 
in which 234 spurious wavenumber values were omitted, and the 125 new
experimental data are included. Thus, 
new experimental data obtained in the present work are in good agreement 
with the wavenumbers of all other lines referencing 
the initial and final rovibronic levels of those questionable lines.

\section*{Results}

Optimal values of 1050 triplet rovibronic energy levels
of $D_2$ have been obtained by the minimization of Eq.(\ref{neviazka}) 
with the input experimental 
data set consisting of 3713 wavenumber values (3588 old and 125 new).
The unknown shift between levels of ortho- and para- deuterium was
obtained by least squares analysis of the $a^3\Sigma_g^+$, $v = 0$,
$N = 0$ - 18 levels with odd and even values of the rotational
quantum number N. All the energy levels were obtained relative to the
lowest vibro-rotational level ($v = 0$, $N=0$) of 
the $a^3\Sigma_g^+$ state  
and presented in Table \ref{NewLevels} together with the SD of the
semi-empirical determination. Absolute values of the triplet 
rovibronic levels relative to the $X^1\Sigma_g^+$, $v = 0$, $N = 0$ ground
rovibronic state of $D_2$ may be obtained by adding the difference
$(E_{a00} - E_{X00})=95348,22$ cm $^{-1}$ from Ref. \cite{Crosswhite} to our 
level values.   
 
Energy level values obtained in the present work are now incorporated 
into the database of the Atomic and Molecular (A+M) Data Unit of the 
International Atomic Energy Agency (IAEA).

\section*{Conclusion}


A detailed comparison of the optimal set of rovibronic levels with
the data reported in Ref. \cite{Nist,Crosswhite} will be presented in 
a separate paper (see also \cite{LU2008xxx}).  

Most of
the differences between our data and those from Ref. \cite{Crosswhite}
are less than $0.05$ cm$^{-1}$, the values reported in 
Ref.\cite{Crosswhite} as an
estimation of the one standard deviation uncertainty of the energy
levels. On the other hand, most of those differences are
sufficiently larger than the SD of our data, which normally
are within the range of $0.004 \div 0.03$ cm$^{-1}$ and depend on
the value of the rotational and vibrational quantum numbers. 
Therefore, the deviations of the data obtained
in Ref.\cite{Crosswhite} from our optimal level values are
significant.

Thus, the optimizational approach
to the problem  allows us to obtain
significantly higher precision in derivation of the energy level
values from measured wavenumbers of rovibronic spectral lines of
$D_2$ molecule.

\section*{Acknowledgments}

The authors are indebted to Prof. S. Ross for providing  
the electronic version of the wavelength tables for lines of 
the $D_2$ emission
spectrum measured and assigned by G.H. Dieke and co-workers
(Appendix C. from Ref.\cite{Crosswhite}), and to Mr. M.S. Ryazanov for
development of the computer code and helpful discussions. We also
acknowledge financial support from the Russian Foundation for
Basic Research (the Grant \#06-03-32663a). 
BPL is thankful to Dr. R.E.H. Clark, Dr. A. Nichols and 
Mr. D. Humbert for useful discussions, and to
the A+M Data Unit of IAEA for financial support of his visits and 
hospitality.
\thispagestyle{empty}

{\def\baselinestretch{0.8}
\footnotesize


\newpage

\begin{table}
\caption{\label{tabband}
RMS uncertainties $\sigma$ of experimental wavenumbers of rovibronic lines for 
the (v'-v") bands of various electronic transitions of $D_2$  molecule
obtained in the present work by pure statistical analysis independent
from the estimates reported in original experimental papers. Standard deviations
for the uncertainties are shown in brackets in units of last significant digit.}

\begin{tabular}{|c|rcr|l|c|rcr|l|}

\hline
Electron transition & \multicolumn{3}{c|}{Band} & $\sigma$, cm$^{-1}$ & Electron transition & \multicolumn{3}{c|}{Band} & $\sigma$, cm$^{-1}$ \\ \hline  
$c^3\Pi_u-a^3\Sigma_g^+$      &$ 3 $&-&$ 0 $&$ 0.076(7)  $&   $d^3\Pi_u-a^3\Sigma_g^+$        &$ 7 $&-&$ 5 $&$ 0.0245(5) $\\ \cline{1-5} 
$a^3\Sigma_g^+-c^3\Pi_u$      &$ 3 $&-&$ 0 $&$ 0.076(7)  $&                                   &$ 7 $&-&$ 6 $&$ 0.0072(3) $\\             
                              &$ 4 $&-&$ 1 $&$ 0.052(3)  $&                                   &$ 7 $&-&$ 7 $&$ 0.0245(5) $\\             
                              &$ 4 $&-&$ 2 $&$ 0.0245(5) $&                                   &$ 7 $&-&$ 8 $&$ 0.052(3)  $\\             
                              &$ 5 $&-&$ 3 $&$ 0.052(3)  $&                                   &$ 7 $&-&$ 9 $&$ 0.052(3)  $\\ \cline{1-5} 
$d^3\Pi_u-a^3\Sigma_g^+$      &$ 0 $&-&$ 0 $&$ 0.0154(3) $&                                   &$ 8 $&-&$ 5 $&$ 0.09(1)   $\\
                              &$ 0 $&-&$ 1 $&$ 0.0154(3) $&                                   &$ 8 $&-&$ 6 $&$ 0.0072(3) $\\
                              &$ 0 $&-&$ 2 $&$ 0.052(3)  $&                                   &$ 8 $&-&$ 7 $&$ 0.0245(5) $\\             
                              &$ 1 $&-&$ 0 $&$ 0.0154(3) $&                                   &$ 8 $&-&$ 8 $&$ 0.12(3)   $\\             
                              &$ 1 $&-&$ 1 $&$ 0.0154(3) $&                                   &$ 8 $&-&$ 9 $&$ 0.0072(3) $\\             
                              &$ 1 $&-&$ 2 $&$ 0.0154(3) $&                                   &$ 8 $&-&$ 10$&$ 0.052(3)  $\\             
                              &$ 1 $&-&$ 3 $&$ 0.0072(3) $&                                   &$ 9 $&-&$ 7 $&$ 0.10(2)   $\\             
                              &$ 2 $&-&$ 0 $&$ 0.0154(3) $&                                   &$ 9 $&-&$ 8 $&$ 0.09(1)   $\\             
                              &$ 2 $&-&$ 1 $&$ 0.0072(3) $&                                   &$ 9 $&-&$ 9 $&$ 0.052(3)  $\\             
                              &$ 2 $&-&$ 2 $&$ 0.0154(3) $&                                   &$ 9 $&-&$ 10$&$ 0.052(3)  $\\             
                              &$ 2 $&-&$ 3 $&$ 0.0154(3) $&                                   &$ 10$&-&$ 8 $&$ 0.052(3)  $\\             
                              &$ 2 $&-&$ 4 $&$ 0.0154(3) $&                                   &$ 10$&-&$ 9 $&$ 0.052(3)  $\\ \cline{6-10}
                              &$ 3 $&-&$ 1 $&$ 0.035(1)  $&   $e^3\Sigma_u^+-a^3\Sigma_g^+$   &$ 0 $&-&$ 0 $&$ 0.0154(3) $\\             
                              &$ 3 $&-&$ 2 $&$ 0.0154(3) $&                                   &$ 0 $&-&$ 1 $&$ 0.0154(3) $\\             
                              &$ 3 $&-&$ 3 $&$ 0.0154(3) $&                                   &$ 1 $&-&$ 0 $&$ 0.0154(3) $\\             
                              &$ 3 $&-&$ 4 $&$ 0.0245(5) $&                                   &$ 1 $&-&$ 1 $&$ 0.0072(3) $\\             
                              &$ 3 $&-&$ 5 $&$ 0.0154(3) $&                                   &$ 1 $&-&$ 2 $&$ 0.035(1)  $\\             
                              &$ 4 $&-&$ 2 $&$ 0.0154(3) $&                                   &$ 1 $&-&$ 4 $&$ 0.052(3)  $\\             
                              &$ 4 $&-&$ 3 $&$ 0.0154(3) $&                                   &$ 2 $&-&$ 0 $&$ 0.035(1)  $\\             
                              &$ 4 $&-&$ 4 $&$ 0.0245(5) $&                                   &$ 2 $&-&$ 1 $&$ 0.0154(3) $\\             
                              &$ 4 $&-&$ 5 $&$ 0.0154(3) $&                                   &$ 2 $&-&$ 2 $&$ 0.0245(5) $\\             
                              &$ 4 $&-&$ 6 $&$ 0.044(2)  $&                                   &$ 2 $&-&$ 3 $&$ 0.035(1)  $\\             
                              &$ 5 $&-&$ 2 $&$ 0.0245(5) $&                                   &$ 3 $&-&$ 0 $&$ 0.0154(3) $\\             
                              &$ 5 $&-&$ 3 $&$ 0.0072(3) $&                                   &$ 3 $&-&$ 1 $&$ 0.0072(3) $\\             
                              &$ 5 $&-&$ 4 $&$ 0.0154(3) $&                                   &$ 3 $&-&$ 2 $&$ 0.0072(3) $\\             
                              &$ 5 $&-&$ 5 $&$ 0.0245(5) $&                                   &$ 3 $&-&$ 3 $&$ 0.0154(3) $\\             
                              &$ 5 $&-&$ 6 $&$ 0.0072(3) $&                                   &$ 3 $&-&$ 4 $&$ 0.0245(5) $\\             
                              &$ 5 $&-&$ 7 $&$ 0.035(1)  $&                                   &$ 3 $&-&$ 6 $&$ 0.063(5)  $\\             
                              &$ 5 $&-&$ 8 $&$ 0.12(3)   $&                                   &$ 3 $&-&$ 7 $&$ 0.09(1)   $\\             
                              &$ 6 $&-&$ 3 $&$ 0.0154(3) $&                                   &$ 4 $&-&$ 0 $&$ 0.0072(3) $\\             
                              &$ 6 $&-&$ 4 $&$ 0.0245(5) $&                                   &$ 4 $&-&$ 1 $&$ 0.0154(3) $\\             
                              &$ 6 $&-&$ 5 $&$ 0.0072(3) $&                                   &$ 4 $&-&$ 2 $&$ 0.0154(3) $\\             
                              &$ 6 $&-&$ 6 $&$ 0.0154(3) $&                                   &$ 4 $&-&$ 3 $&$ 0.0072(3) $\\             
                              &$ 6 $&-&$ 7 $&$ 0.0072(3) $&                                   &$ 4 $&-&$ 4 $&$ 0.0154(3) $\\             
                              &$ 6 $&-&$ 8 $&$ 0.0072(3) $&                                   &$ 4 $&-&$ 5 $&$ 0.0154(3) $\\                                               
                              &$ 7 $&-&$ 4 $&$ 0.0245(5) $&                                   &$ 4 $&-&$ 7 $&$ 0.035(1)  $\\ \hline                                        
\end{tabular}                                                                                 
\end{table}                                                                                   

\newpage                                                                                      

\begin{table}

\begin{tabular}{|c|rcr|l|c|rcr|l|}
\multicolumn{10}{l}{\bf Table \ref{tabband}. \rm (Continued.)} \\
\hline
Electron transition & \multicolumn{3}{c|}{Band} & $\sigma$, cm$^{-1}$ & Electron transition & \multicolumn{3}{c|}{Band} & $\sigma$, cm$^{-1}$ \\ \hline  
$e^3\Sigma_u^+-a^3\Sigma_g^+$ &$ 4 $&-&$ 8 $&$ 0.052(3)  $& $f^3\Sigma_u^+-a^3\Sigma_g^+$     &$ 1 $&-&$ 0 $&$ 0.0154(3) $\\
                              &$ 5 $&-&$ 1 $&$ 0.035(1)  $&                                   &$ 1 $&-&$ 1 $&$ 0.0154(3) $\\
                              &$ 5 $&-&$ 2 $&$ 0.0154(3) $&                                   &$ 1 $&-&$ 2 $&$ 0.0245(5) $\\
                              &$ 5 $&-&$ 3 $&$ 0.0072(3) $&                                   &$ 2 $&-&$ 0 $&$ 0.0245(5) $\\             
                              &$ 5 $&-&$ 4 $&$ 0.0072(3) $&                                   &$ 2 $&-&$ 1 $&$ 0.0245(5) $\\             
                              &$ 5 $&-&$ 6 $&$ 0.0154(3) $&                                   &$ 2 $&-&$ 2 $&$ 0.0154(3) $\\             
                              &$ 5 $&-&$ 8 $&$ 0.063(5)  $&                                   &$ 2 $&-&$ 3 $&$ 0.0154(3) $\\ \cline{6-10}
                              &$ 6 $&-&$ 1 $&$ 0.0245(5) $& $g^3\Sigma_g^+-c^3\Pi_u$          &$ 0 $&-&$ 0 $&$ 0.0245(5) $\\             
                              &$ 6 $&-&$ 2 $&$ 0.0245(5) $&                                   &$ 0 $&-&$ 1 $&$ 0.063(5)  $\\             
                              &$ 6 $&-&$ 3 $&$ 0.0154(3) $&                                   &$ 1 $&-&$ 0 $&$ 0.076(7)  $\\             
                              &$ 6 $&-&$ 4 $&$ 0.0154(3) $&                                   &$ 1 $&-&$ 1 $&$ 0.0154(3) $\\             
                              &$ 6 $&-&$ 5 $&$ 0.0072(3) $&                                   &$ 1 $&-&$ 2 $&$ 0.035(1)  $\\             
                              &$ 6 $&-&$ 6 $&$ 0.0245(5) $&                                   &$ 2 $&-&$ 1 $&$ 0.044(2)  $\\             
                              &$ 7 $&-&$ 2 $&$ 0.0154(3) $&                                   &$ 2 $&-&$ 2 $&$ 0.035(1)  $\\             
                              &$ 7 $&-&$ 3 $&$ 0.0154(3) $&                                   &$ 3 $&-&$ 2 $&$ 0.0154(3) $\\             
                              &$ 7 $&-&$ 4 $&$ 0.0072(3) $&                                   &$ 3 $&-&$ 3 $&$ 0.0154(3) $\\             
                              &$ 7 $&-&$ 5 $&$ 0.0154(3) $&                                   &$ 4 $&-&$ 3 $&$ 0.035(1)  $\\             
                              &$ 7 $&-&$ 6 $&$ 0.044(2)  $&                                   &$ 4 $&-&$ 4 $&$ 0.0154(3) $\\ \cline{6-10}
                              &$ 7 $&-&$ 7 $&$ 0.0245(5) $& $g^3\Sigma_g^+-e^3\Sigma_u^+$     &$ 0 $&-&$ 0 $&$ 0.044(2)  $\\             
                              &$ 8 $&-&$ 2 $&$ 0.0245(5) $& 		                      &$ 0 $&-&$ 1 $&$ 0.052(3)  $\\             
                              &$ 8 $&-&$ 3 $&$ 0.0154(3) $& 				      &$ 1 $&-&$ 1 $&$ 0.0154(3) $\\             
                              &$ 8 $&-&$ 4 $&$ 0.0154(3) $& 				      &$ 2 $&-&$ 2 $&$ 0.0072(3) $\\ \cline{6-10}
                              &$ 8 $&-&$ 5 $&$ 0.0154(3) $& $h^3\Sigma_u^+-c^3\Pi_u$          &$ 2 $&-&$ 2 $&$ 0.0154(3) $\\             
                              &$ 8 $&-&$ 6 $&$ 0.0154(3) $&                                   &$ 3 $&-&$ 3 $&$ 0.0072(3) $\\             
                              &$ 8 $&-&$ 7 $&$ 0.0245(5) $&                                   &$ 4 $&-&$ 4 $&$ 0.0072(3) $\\             
                              &$ 8 $&-&$ 8 $&$ 0.0245(5) $&                                   &$ 4 $&-&$ 5 $&$ 0.035(1)  $\\ \cline{6-10}
                              &$ 9 $&-&$ 3 $&$ 0.035(1)  $& $i^3\Pi_g^+-c^3\Pi_u$             &$ 0 $&-&$ 0 $&$ 0.0245(5) $\\             
                              &$ 9 $&-&$ 4 $&$ 0.0072(3) $&                                   &$ 0 $&-&$ 1 $&$ 0.076(7)  $\\             
                              &$ 9 $&-&$ 5 $&$ 0.0154(3) $&                                   &$ 1 $&-&$ 0 $&$ 0.044(2)  $\\             
                              &$ 9 $&-&$ 6 $&$ 0.0154(3) $&                                   &$ 1 $&-&$ 1 $&$ 0.035(1)  $\\             
                              &$ 9 $&-&$ 7 $&$ 0.044(2)  $&                                   &$ 1 $&-&$ 2 $&$ 0.044(2)  $\\             
                              &$ 9 $&-&$ 8 $&$ 0.035(1)  $&                                   &$ 2 $&-&$ 2 $&$ 0.044(2)  $\\             
                              &$ 10$&-&$ 3 $&$ 0.035(1)  $&                                   &$ 3 $&-&$ 3 $&$ 0.0245(5) $\\             
                              &$ 10$&-&$ 4 $&$ 0.035(1)  $&                                   &$ 4 $&-&$ 4 $&$ 0.035(1)  $\\ \cline{6-10}
                              &$ 10$&-&$ 5 $&$ 0.035(1)  $& $i^3\Pi_g^--c^3\Pi_u$             &$ 0 $&-&$ 0 $&$ 0.0245(5) $\\             
                              &$ 10$&-&$ 6 $&$ 0.035(1)  $&                                   &$ 0 $&-&$ 1 $&$ 0.076(7)  $\\             
                              &$ 10$&-&$ 7 $&$ 0.035(1)  $&                                   &$ 1 $&-&$ 0 $&$ 0.044(2)  $\\             
                              &$ 10$&-&$ 8 $&$ 0.035(1)  $&                                   &$ 1 $&-&$ 1 $&$ 0.035(1)  $\\             
                              &$ 10$&-&$ 9 $&$ 0.035(1)  $&                                   &$ 1 $&-&$ 2 $&$ 0.044(2)  $\\ \cline{1-5}             
$f^3\Sigma_u^+-a^3\Sigma_g^+$ &$ 0 $&-&$ 0 $&$ 0.035(1)  $&                                   &$ 2 $&-&$ 1 $&$ 0.0154(3) $\\             
                              &$ 0 $&-&$ 1 $&$ 0.044(2)  $&                                   &$ 2 $&-&$ 2 $&$ 0.044(2)  $\\ \hline      
\end{tabular}                                               
\end{table}          	                                    

\newpage

\begin{table}
\begin{tabular}{|c|rcr|l|c|rcr|l|}
\multicolumn{10}{l}{\bf Table \ref{tabband}. \rm (Continued.)} \\
\hline
Electron transition & \multicolumn{3}{c|}{Band} & $\sigma$, cm$^{-1}$ & Electron transition & \multicolumn{3}{c|}{Band} & $\sigma$, cm$^{-1}$ \\ \hline  
$i^3\Pi_g^--c^3\Pi_u$         &$ 2 $&-&$ 3 $&$ 0.0245(5) $&  $k^3\Pi_u-a^3\Sigma_g^+$       &$ 5 $&-&$ 5 $&$ 0.035(1)  $\\  
                              &$ 3 $&-&$ 3 $&$ 0.0245(5) $&                                 &$ 5 $&-&$ 6 $&$ 0.035(1)  $\\  
                              &$ 4 $&-&$ 4 $&$ 0.035(1)  $&                                 &$ 5 $&-&$ 7 $&$ 0.052(3)  $\\ \cline{1-5}
$i^3\Pi_g-e^3\Sigma_u^+$      &$ 0 $&-&$ 0 $&$ 0.035(1)  $&                                 &$ 6 $&-&$ 4 $&$ 0.035(1)  $\\  
                              &$ 1 $&-&$ 1 $&$ 0.035(1)  $&                                 &$ 6 $&-&$ 5 $&$ 0.0154(3) $\\  
                              &$ 2 $&-&$ 2 $&$ 0.076(7)  $&  $n^3\Pi_u-a^3\Sigma_g^+$       &$ 0 $&-&$ 0 $&$ 0.044(2)  $\\ \cline{1-5}
$j^3\Delta_g^+-c^3\Pi_u$      &$ 0 $&-&$ 0 $&$ 0.0245(5) $&                                 &$ 0 $&-&$ 1 $&$ 0.052(3)  $\\  
                              &$ 1 $&-&$ 0 $&$ 0.044(2)  $&                                 &$ 1 $&-&$ 0 $&$ 0.0245(5) $\\  
                              &$ 1 $&-&$ 1 $&$ 0.0245(5) $&                                 &$ 1 $&-&$ 1 $&$ 0.035(1)  $\\                   
                              &$ 2 $&-&$ 1 $&$ 0.035(1)  $&                                 &$ 1 $&-&$ 2 $&$ 0.044(2)  $\\                    
                              &$ 2 $&-&$ 2 $&$ 0.0245(5) $&                                 &$ 2 $&-&$ 1 $&$ 0.10(2)   $\\                    
                              &$ 3 $&-&$ 3 $&$ 0.0154(3) $&                                 &$ 2 $&-&$ 2 $&$ 0.044(2)  $\\ \cline{1-5}
$j^3\Delta_g^--c^3\Pi_u$      &$ 0 $&-&$ 0 $&$ 0.0245(5) $&                                 &$ 2 $&-&$ 3 $&$ 0.0154(3) $\\                    
                              &$ 0 $&-&$ 1 $&$ 0.0154(3) $&                                 &$ 3 $&-&$ 2 $&$ 0.063(5)  $\\                    
                              &$ 1 $&-&$ 0 $&$ 0.044(2)  $&                                 &$ 3 $&-&$ 3 $&$ 0.044(2)  $\\                    
                              &$ 1 $&-&$ 1 $&$ 0.0245(5) $&                                 &$ 3 $&-&$ 4 $&$ 0.0072(3) $\\ \cline{6-10}       
                              &$ 1 $&-&$ 2 $&$ 0.0245(5) $&   $p^3\Sigma_g^+-c^3\Pi_u$      &$ 0 $&-&$ 0 $&$ 0.076(7)  $\\
                              &$ 2 $&-&$ 1 $&$ 0.035(1)  $&                                 &$ 0 $&-&$ 1 $&$ 0.13(3)   $\\
                              &$ 2 $&-&$ 2 $&$ 0.0245(5) $&                                 &$ 1 $&-&$ 0 $&$ 0.09(1)   $\\                            
                              &$ 2 $&-&$ 3 $&$ 0.0072(3) $&                                 &$ 1 $&-&$ 1 $&$ 0.0154(3) $\\                         
                              &$ 3 $&-&$ 2 $&$ 0.052(3)  $&                                 &$ 1 $&-&$ 2 $&$ 0.09(1)   $\\                         
                              &$ 3 $&-&$ 3 $&$ 0.0154(3) $&                                 &$ 2 $&-&$ 1 $&$ 0.044(2)  $\\                         
                              &$ 3 $&-&$ 4 $&$ 0.035(1)  $&                                 &$ 2 $&-&$ 2 $&$ 0.063(5)  $\\                         
                              &$ 4 $&-&$ 4 $&$ 0.052(3)  $&                                 &$ 2 $&-&$ 3 $&$ 0.12(3)   $\\                    
                              &$ 5 $&-&$ 5 $&$ 0.076(7)  $&                                 &$ 3 $&-&$ 3 $&$ 0.0154(3) $\\ \cline{1-5}
$j^3\Delta_g-e^3\Sigma_u^+$   &$ 0 $&-&$ 0 $&$ 0.10(2)   $&                                 &$ 4 $&-&$ 4 $&$ 0.0245(5) $\\ \hline
$k^3\Pi_u-a^3\Sigma_g^+$      &$ 0 $&-&$ 0 $&$ 0.0245(5) $&   $p^3\Sigma_g^+-e^3\Sigma_u^+$ &$ 0 $&-&$ 0 $&$ 0.063(5)  $\\                         
                              &$ 0 $&-&$ 1 $&$ 0.0245(5) $&                                 &$ 0 $&-&$ 1 $&$ 0.063(5)  $\\                         
                              &$ 1 $&-&$ 0 $&$ 0.035(1)  $&                                 &$ 1 $&-&$ 1 $&$ 0.0245(5) $\\                         
                              &$ 1 $&-&$ 1 $&$ 0.035(1)  $&                                 &$ 2 $&-&$ 2 $&$ 0.052(3)  $\\ \cline{6-10}            
                              &$ 1 $&-&$ 2 $&$ 0.035(1)  $&   $q^3\Sigma_g^+-c^3\Pi_u$      &$ 0 $&-&$ 0 $&$ 0.0245(5) $\\ \cline{6-10}            
                              &$ 2 $&-&$ 1 $&$ 0.0245(5) $&   $r^3\Pi_g^+-c^3\Pi_u$         &$ 0 $&-&$ 0 $&$ 0.052(3)  $\\ \cline{6-10}            
                              &$ 2 $&-&$ 2 $&$ 0.0154(3) $&   $r^3\Pi_g^--c^3\Pi_u$         &$ 0 $&-&$ 0 $&$ 0.052(3)  $\\ \cline{6-10}            
                              &$ 2 $&-&$ 3 $&$ 0.0072(3) $&   $r^3\Pi_g-e^3\Sigma_u^+$      &$ 0 $&-&$ 0 $&$ 0.09(2)   $\\ \cline{6-10}             
                              &$ 3 $&-&$ 2 $&$ 0.0154(3) $&   $s^3\Delta_g^--c^3\Pi_u$      &$ 0 $&-&$ 0 $&$ 0.035(1)  $\\ \cline{6-10}            
                              &$ 3 $&-&$ 3 $&$ 0.0245(5) $&   $u^3\Pi_u-a^3\Sigma_g^+$      &$ 1 $&-&$ 0 $&$ 0.044(2)  $\\                         
                              &$ 3 $&-&$ 4 $&$ 0.0245(5) $&                                 &$ 1 $&-&$ 1 $&$ 0.0154(3) $\\                         
                              &$ 4 $&-&$ 3 $&$ 0.0245(5) $&                                 &$ 1 $&-&$ 2 $&$ 0.063(5)  $\\
                              &$ 4 $&-&$ 4 $&$ 0.044(2)  $&                                 &$ 2 $&-&$ 1 $&$ 0.13(3)   $\\                         
                              &$ 4 $&-&$ 5 $&$ 0.0245(5) $&                                 &$ 2 $&-&$ 2 $&$ 0.0245(5) $\\                         
                              &$ 5 $&-&$ 4 $&$ 0.052(3)  $&                                 &$ 2 $&-&$ 3 $&$ 0.10(2)   $\\ \hline                  
\end{tabular}
\end{table}


\newpage

\begin{table}

\caption{\label{dellines}
Upper and lower rovibronic levels of spectral lines, which wavenumber 
values (reported in various articles) were excluded as outliers 
from further statistical analysis. $\sigma_{\nu}$ is experimental 
RMS uncertainty for all other lines of the same band.
O-C denotes the differences between Observed wavenumber values and 
those Calculated as differences of corresponding optimal energy level 
values obtained in the present work.
}

\begin{tabular}{|c|c|c|c|c|c|c|c|c|}
\hline
Upper electronic state & v' & N' & Lower electronic state & v'' & N'' & $\sigma_{\nu}$ & O-C & Source \\ \hline
$a^3\Sigma_g^+$ & 3  & 2  & $c^3\Pi_u^+   $ & 2 & 3  & 0.007 & +0.02 & \cite{DabrHerz}   \\
$a^3\Sigma_g^+$ & 3  & 4  & $c^3\Pi_u^+   $ & 2 & 3  & 0.007 & +0.87 & \cite{DabrHerz}   \\
$h^3\Sigma_g^+$ & 3  & 3  & $c^3\Pi_u^+   $ & 3 & 4  & 0.007 & -0.09 & \cite{Crosswhite} \\
$e^3\Sigma_u^+$ & 7  & 6  & $a^3\Sigma_g^+$ & 3 & 7  & 0.007 & -0.14 & \cite{Dieke}      \\
$e^3\Sigma_u^+$ & 7  & 8  & $a^3\Sigma_g^+$ & 3 & 7  & 0.007 & -0.02 & \cite{Dieke}      \\
$e^3\Sigma_u^+$ & 3  & 7  & $a^3\Sigma_g^+$ & 1 & 8  & 0.007 & -0.05 & \cite{Crosswhite} \\
$e^3\Sigma_u^+$ & 4  & 3  & $a^3\Sigma_g^+$ & 3 & 2  & 0.007 & -0.02 & \cite{Crosswhite} \\
\hline
\end{tabular}
\end{table}

\newpage

\begin{table}
\begin{tabular}{|c|c|c|c|c|c|c|c|c|}
\multicolumn{9}{l}{\bf Table \ref{dellines}. \rm (Continued.)} \\
\hline               
Upper electronic state & v' & N' & Lower electronic state & v'' & N'' & $\sigma_{\nu}$ & O-C & Source   \\ \hline
$e^3\Sigma_u^+$ & 2  & 6  & $a^3\Sigma_g^+$ & 0 & 7  & 0.015 & +0.26 & \cite{Dieke}      \\
$e^3\Sigma_u^+$ & 2  & 8  & $a^3\Sigma_g^+$ & 0 & 7  & 0.015 & -0.05 & \cite{Dieke}      \\
$e^3\Sigma_u^+$ & 2  & 9  & $a^3\Sigma_g^+$ & 0 & 8  & 0.015 & +0.05 & \cite{Dieke}      \\
$e^3\Sigma_u^+$ & 4  & 5  & $a^3\Sigma_g^+$ & 2 & 6  & 0.015 & -0.13 & \cite{Dieke}      \\
$e^3\Sigma_u^+$ & 4  & 7  & $a^3\Sigma_g^+$ & 2 & 8  & 0.015 & +0.06 & \cite{Dieke}      \\
$e^3\Sigma_u^+$ & 6  & 5  & $a^3\Sigma_g^+$ & 4 & 6  & 0.015 & -0.06 & \cite{Dieke}      \\
\hline
\end{tabular}
\end{table}

\newpage

\begin{table}
\begin{tabular}{|c|c|c|c|c|c|c|c|c|}
\multicolumn{9}{l}{\bf Table \ref{dellines}. \rm (Continued.)} \\
\hline               
Upper electronic state & v' & N' & Lower electronic state & v'' & N'' & $\sigma_{\nu}$ & O-C & Source   \\ \hline
$d^3\Pi_u^+   $ & 2  & 1  & $a^3\Sigma_g^+$ & 2 & 2  & 0.015 & -0.08 & \cite{DiekeBlue}  \\
$d^3\Pi_u^+   $ & 2  & 6  & $a^3\Sigma_g^+$ & 2 & 5  & 0.015 & +1.33 & \cite{DiekeBlue}  \\
$d^3\Pi_u^+   $ & 2  & 7  & $a^3\Sigma_g^+$ & 2 & 6  & 0.015 & -1.20 & \cite{DiekeBlue}  \\
$d^3\Pi_u^+   $ & 2  & 7  & $a^3\Sigma_g^+$ & 2 & 8  & 0.015 & +0.06 & \cite{DiekeBlue}  \\
$d^3\Pi_u^+   $ & 3  & 4  & $a^3\Sigma_g^+$ & 2 & 5  & 0.015 & -0.50 & \cite{DiekeBlue}  \\
$d^3\Pi_u^+   $ & 3  & 5  & $a^3\Sigma_g^+$ & 2 & 6  & 0.015 & +0.31 & \cite{DiekeBlue}  \\
\hline
\end{tabular}
\end{table}

\newpage

\begin{table}
\begin{tabular}{|c|c|c|c|c|c|c|c|c|}
\multicolumn{9}{l}{\bf Table \ref{dellines}. \rm (Continued.)} \\
\hline               
Upper electronic state & v' & N' & Lower electronic state & v'' & N'' & $\sigma_{\nu}$ & O-C & Source   \\ \hline
$g^3\Sigma_g^+$ & 1  & 3  & $c^3\Pi_u^+   $ & 1 & 4  & 0.015 & +0.12 & \cite{Crosswhite} \\
$g^3\Sigma_g^+$ & 1  & 6  & $c^3\Pi_u^+   $ & 1 & 7  & 0.015 & +0.16 & \cite{Crosswhite} \\
$g^3\Sigma_g^+$ & 3  & 2  & $c^3\Pi_u^-   $ & 2 & 2  & 0.015 & -0.05 & \cite{Crosswhite} \\
$g^3\Sigma_g^+$ & 3  & 3  & $c^3\Pi_u^-   $ & 2 & 3  & 0.015 & -0.04 & \cite{Crosswhite} \\
$g^3\Sigma_g^+$ & 3  & 3  & $c^3\Pi_u^+   $ & 2 & 4  & 0.015 & +0.08 & \cite{Crosswhite} \\
$g^3\Sigma_g^+$ & 3  & 3  & $c^3\Pi_u^+   $ & 3 & 2  & 0.015 & -0.08 & \cite{Crosswhite} \\
\hline
\end{tabular}
\end{table}

\newpage

\begin{table}
\begin{tabular}{|c|c|c|c|c|c|c|c|c|}
\multicolumn{9}{l}{\bf Table \ref{dellines}. \rm (Continued.)} \\
\hline               
Upper electronic state & v' & N' & Lower electronic state & v'' & N'' & $\sigma_{\nu}$ & O-C & Source   \\ \hline
$k^3\Pi_u^+   $ & 0  & 7  & $a^3\Sigma_g^+$ & 0 & 6  & 0.024 & +0.17 & \cite{Crosswhite} \\
$k^3\Pi_u^+   $ & 0  & 7  & $a^3\Sigma_g^+$ & 0 & 8  & 0.024 & -0.01 & \cite{Crosswhite} \\
$k^3\Pi_u^-   $ & 0  & 8  & $a^3\Sigma_g^+$ & 0 & 8  & 0.024 & +0.16 & \cite{Crosswhite} \\
$k^3\Pi_u^+   $ & 0  & 4  & $a^3\Sigma_g^+$ & 1 & 3  & 0.024 & +0.19 & \cite{Crosswhite} \\
$k^3\Pi_u^-   $ & 0  & 8  & $a^3\Sigma_g^+$ & 1 & 8  & 0.024 & -0.04 & \cite{Crosswhite} \\
$k^3\Pi_u^+   $ & 2  & 3  & $a^3\Sigma_g^+$ & 1 & 4  & 0.024 & -0.10 & \cite{Crosswhite} \\
\hline
\end{tabular}
\end{table}

\newpage

\begin{table}
\begin{tabular}{|c|c|c|c|c|c|c|c|c|}
\multicolumn{9}{l}{\bf Table \ref{dellines}. \rm (Continued.)} \\
\hline               
Upper electronic state & v' & N' & Lower electronic state & v'' & N'' & $\sigma_{\nu}$ & O-C & Source   \\ \hline
$e^3\Sigma_u^+$ & 1  & 8  & $a^3\Sigma_g^+$ & 2 & 9  & 0.034 & +0.17 & \cite{Crosswhite} \\
$e^3\Sigma_u^+$ & 2  & 10 & $a^3\Sigma_g^+$ & 0 & 11 & 0.034 & -0.14 & \cite{Crosswhite} \\
$e^3\Sigma_u^+$ & 10 & 3  & $a^3\Sigma_g^+$ & 4 & 4  & 0.034 & +0.25 & \cite{Crosswhite} \\
$e^3\Sigma_u^+$ & 10 & 4  & $a^3\Sigma_g^+$ & 6 & 3  & 0.034 & -0.09 & \cite{Crosswhite} \\
$d^3\Pi_u^+   $ & 2  & 5  & $a^3\Sigma_g^+$ & 1 & 4  & 0.034 & -0.13 & \cite{DiekeBlue}  \\
$d^3\Pi_u^+   $ & 4  & 1  & $a^3\Sigma_g^+$ & 5 & 0  & 0.034 & +1.03 & \cite{DiekeBlue}  \\
\hline
\end{tabular}
\end{table}

\newpage

\begin{table}
\begin{tabular}{|c|c|c|c|c|c|c|c|c|}
\multicolumn{9}{l}{\bf Table \ref{dellines}. \rm (Continued.)} \\
\hline               
Upper electronic state & v' & N' & Lower electronic state & v'' & N'' & $\sigma_{\nu}$ & O-C & Source   \\ \hline
$j^3\Delta_g^+$ & 1  & 2  & $c^3\Pi_u^-   $ & 0 & 2  & 0.044 & +0.14 & \cite{Crosswhite} \\
$j^3\Delta_g^-$ & 1  & 4  & $c^3\Pi_u^+   $ & 0 & 4  & 0.044 & -0.18 & \cite{Crosswhite} \\
$j^3\Delta_g^-$ & 1  & 6  & $c^3\Pi_u^-   $ & 0 & 7  & 0.044 & +0.26 & \cite{Crosswhite} \\
$k^3\Pi_u^-   $ & 5  & 4  & $a^3\Sigma_g^+$ & 4 & 4  & 0.051 & -0.13 & \cite{Crosswhite} \\
$e^3\Sigma_u^+$ & 3  & 0  & $a^3\Sigma_g^+$ & 6 & 1  & 0.063 & -0.19 & \cite{Crosswhite} \\
$g^3\Sigma_g^+$ & 0  & 2  & $c^3\Pi_u^+   $ & 1 & 3  & 0.063 & -0.18 & \cite{Crosswhite} \\
$c^3\Pi_u^+   $ & 3  & 3  & $a^3\Sigma_g^+$ & 0 & 2  & 0.075 & +0.26 & \cite{Crosswhite} \\
$e^3\Sigma_u^+$ & 4  & 2  & $a^3\Sigma_g^+$ & 6 & 1  & 0.26  & -1.39 & \cite{DiekePorto} \\
$e^3\Sigma_u^+$ & 4  & 8  & $a^3\Sigma_g^+$ & 6 & 9  & 0.26  & -3.31 & \cite{DiekePorto} \\
$e^3\Sigma_u^+$ & 4  & 9  & $a^3\Sigma_g^+$ & 6 & 10 & 0.26  & -2.24 & \cite{DiekePorto} \\
$e^3\Sigma_u^+$ & 4  & 11 & $a^3\Sigma_g^+$ & 6 & 10 & 0.26  & +2.24 & \cite{DiekePorto} \\ \hline
\end{tabular}
\end{table}


\newpage

\begin{table}
\caption{\label{NewLines}
Experimental values of the wavenumbers (in cm$^{-1}$) of some spectral lines of
$D_2$ obtained in the present work (P.W.)and reported in \cite{Crosswhite}.
O-C denotes the differences between Observed wavenumber values and those Calculated
as differences of corresponding optimal energy level values obtained in the present work.}

\begin{tabular}{@{}lllrrrr}
\hline
 Electronic transition        &  Band    &  Line   &  \cite{Crosswhite}  &  O-C    &  P.W.              &  O-C    \\
\hline
$e^3\Sigma_u^+-a^3\Sigma_g^+$ & 2-0      & P11     & 13751.08            & -0.14   & 13751.29           &  0.07   \\
                              & 3-0      & R6      & 15868.98            & -0.05   & 15869.01           & -0.02   \\
                              &          & R9      & 15705.21            &  0.23   & 15704.98           &  0.00   \\
                              & 4-1      & P8      & 14862.89            & -0.08   & 14862.95           & -0.02   \\
                              & 5-2      & P6      & 14524.31            &  0.07   & 14524.26           &  0.02   \\	
                              & 6-2      & R4      & 15977.48            & -0.06   & 15977.56           &  0.02   \\
                              & 6-3      & P9      & 13703.95            &  0.09   & 13703.87           &  0.01   \\
                              & 7-3      & P5      & 15107.67            & -0.04   & 15107.71           &  0.00   \\
                              & 8-3      & R0      & 16354.21            &  0.05   & 16354.19           &  0.03   \\
                              & 8-4      & P5      & 14468.77            & -0.10   & 14468.86           & -0.01   \\
                              &          & R3      & 14727.33            & -0.06   & 14727.38           & -0.01   \\
                              & 9-5      & P4      & 13861.27            & -0.19   & 13861.45           & -0.01   \\
                              &          & R4      & 14002.82            & -0.11   & 14002.91           & -0.02   \\
                              & 10-4     & P4      & 16124.14            &  0.25   & 16123.94           &  0.05   \\
$f^3\Sigma_u^+-a^3\Sigma_g^+$ & 1-1      & R2      & 20346.49            &  0.07   & 20346.42           &  0.00   \\
                              & 1-2      & R0      & 18564.54            &  0.21   & 18564.19           & -0.14   \\
                              & 2-1      & P3      & 21653.13            &  0.20   & 21653.04           &  0.11   \\
                              & 2-3      & P3      & 18242.42            &  0.09   & 18242.40           &  0.07   \\
$d^3\Pi_u-a^3\Sigma_g^+$      & 0-1      & R9      & 15019.54            &  0.10   & 15019.44           &  0.00   \\
                              & 1-0      & P2      & 18207.70            &  0.10   & 18207.60           &  0.00   \\
                              &          & P5      & 18065.38            &  0.06   & 18065.36           &  0.04   \\
                              & 1-1      & Q1      & 16460.87            & -0.07   & 16460.93           & -0.01   \\
                              &          & Q10     & 16267.60            & -0.05   & 16267.64           & -0.01   \\
                              & 1-2      & Q10     & 14582.86            &  0.03   & 14582.84           &  0.01   \\
                              & 2-0      & Q1      & 19823.53            &  0.09   & 19823.45           &  0.01   \\
                              & 2-1      & R1      & 18065.70            &  0.07   & 18065.68           &  0.05   \\
                              &          & R5      & 18109.66            &  0.04   & 18109.64           &  0.02   \\
                              &          & R6      & 18107.89            & -0.04   & 18107.97           &  0.04   \\
                              &          & R8      & 18089.06            & -0.05   & 18089.10           & -0.01   \\
                              & 2-2      & R7      & 16386.78            &  0.05   & 16386.78           &  0.05   \\
                              &          & R8      & 16383.63            &  0.05   & 16383.58           &  0.00   \\
                              & 2-3      & Q6      & 14538.55            &  0.05   & 14538.51           &  0.01   \\
                              &          & Q9      & 14481.26            &  0.09   & 14481.19           &  0.02   \\
                              & 3-1      & R4      & 19576.76            &  0.17   & 19576.61           &  0.02   \\
                              & 3-2      & R4      & 17842.03            & -0.08   & 17842.10           & -0.01   \\
                              & 3-3      & R5      & 16180.93            & -0.10   & 16181.08           &  0.05   \\
                              & 3-4      & Q6      & 14415.64            &  0.15   & 14415.48           & -0.01   \\
\hline
\end{tabular}
\end{table}

\newpage

\begin{table}
\begin{tabular}{@{}lllrrrr}
\multicolumn{7}{l}{\bf Table \ref{NewLines}. \rm (Continued.)} \\
\hline
 Electronic transition        &  Band    &  Line   &  \cite{Crosswhite}  &  O-C    &  This work         &  O-C    \\
\hline
$d^3\Pi_u-a^3\Sigma_g^+$      & 4-2      & P8      & 18788.84            & -0.05   & 18788.84           & -0.05   \\
                              & 4-3      & P3      & 17401.87            &  0.11   & 17401.74           & -0.02   \\
                              &          & R6      & 17586.74            &  0.07   & 17586.69           &  0.02   \\
                              & 4-5      & Q7      & 14280.99            &  0.06   & 14280.94           &  0.01   \\
                              & 5-2      & Q3      & 20510.44            & -0.14   & 20510.57           & -0.01   \\
                              & 5-3      & Q6      & 18757.31            & -0.17   & 18757.49           &  0.01   \\
                              &          & Q8      & 18677.56            &  0.08   & 18677.49           &  0.01   \\
                              & 6-4      & Q7      & 18420.01            & -0.11   & 18420.10           & -0.02   \\
                              & 7-6      & Q5      & 16727.50            &  0.18   & 16727.38           &  0.06   \\
$k^3\Pi_u-a^3\Sigma_g^+$      & 0-0      & P8      & 21966.23            & -0.01   & 21966.21           & -0.03   \\
                              &          & Q8      & 22180.52            &  0.16   & 22180.39           &  0.03   \\
                              &          & R6      & 22452.89            &  0.17   & 22452.75           &  0.03   \\
                              & 0-1      & Q8      & 20406.67            & -0.04   & 20406.68           & -0.03   \\
                              &          & R3      & 20613.91            &  0.19   & 20613.78           &  0.06   \\
                              & 1-2      & P4      & 20217.44            &  0.11   & 20217.32           & -0.01   \\
                              & 2-3      & P2      & 20142.72            & -0.14   & 20142.76           & -0.10   \\
                              &          & R2      & 20282.53            &  0.21   & 20282.45           &  0.13   \\
                              &          & R4      & 20320.54            &  0.09   & 20320.48           &  0.03   \\
                              & 3-3      & Q6      & 21592.35            &  0.22   & 21592.11           & -0.02   \\
                              & 4-3      & R4      & 23156.99            &  0.10   & 23156.92           &  0.03   \\
                              & 4-5      & P5      & 19755.38            & -0.15   & 19755.42           & -0.11   \\
                              & 5-4      & Q4      & 22761.98            & -0.13   & 22762.13           &  0.02   \\
$n^3\Pi_u-a^3\Sigma_g^+$      & 1-2      & Q5      & 22881.68            & -0.05   & 22881.71           & -0.02   \\
                              & 2-3      & P3      & 22670.82            & -0.12   & 22670.88           & -0.06   \\
$h^3\Sigma_g^+-c^3\Pi_u$      & 3-3      & P4      & 16779.23            & -0.09   & 16779.32           &  0.00   \\
$g^3\Sigma_g^+-c^3\Pi_u$      & 0-0      & P5      & 16588.93            &  0.13   & 16588.87           &  0.07   \\
                              & 0-1      & P3      & 15064.96            & -0.18   & 15065.27           &  0.13   \\
                              & 1-1      & P4      & 16502.17            &  0.13   & 16502.08           &  0.04   \\
                              & 2-1      & P5      & 17846.80            & -0.27   & 17847.07           &  0.00   \\
                              & 2-2      & P6      & 16164.95            &  0.22   & 16164.69           & -0.04   \\
                              &          & R9      & 16314.06            &  0.01   & 16314.07           &  0.02   \\
                              &          & Q10     & 16045.63            & -0.03   & 16045.64           & -0.02   \\
                              & 3-2      & P4      & 17652.53            &  0.09   & 17652.43           & -0.01   \\
                              &          & Q2      & 17801.97            & -0.06   & 17801.97           & -0.06   \\
                              &          & Q3      & 17765.57            & -0.05   & 17765.67           &  0.05   \\
                              & 3-3      & Q7      & 16024.02            &  0.09   & 16023.98           &  0.05   \\
                              &          & R2      & 16293.82            & -0.08   & 16293.88           & -0.02   \\
                              &          & R6      & 16210.71            & -0.04   & 16210.70           & -0.05   \\
$i^3\Pi_g^+-c^3\Pi_u$         & 0-0      & P11     & 16902.38            & -0.03   & 16902.34           & -0.07   \\
                              &          & Q10     & 17214.20            &  0.13   & 17214.14           &  0.07   \\
\hline
\end{tabular}
\end{table}

\newpage

\begin{table}
\begin{tabular}{@{}lllrrrr}
\multicolumn{7}{l}{\bf Table \ref{NewLines}. \rm (Continued.)} \\
\hline
 Electronic transition        &  Band    &  Line   &  \cite{Crosswhite}  &  O-C    &  This work         &  O-C    \\
\hline
$i^3\Pi_g^+-c^3\Pi_u$         & 1-0      & P7      & 18524.43            &  0.16   & 18524.27           &  0.00   \\
                              &          & Q5      & 18728.49            &  0.20   & 18728.35           &  0.06   \\
                              &          & R1      & 18750.25            & -0.20   & 18750.51           &  0.06   \\
                              &          & R7      & 18957.79            &  0.17   & 18957.66           &  0.04   \\
                              & 2-1      & Q2      & 18425.10            & -0.17   & 18425.19           & -0.08   \\
                              & 3-3      & P8      & 16450.63            & -0.08   & 16450.62           & -0.09   \\
                              &          & R6      & 16849.79            &  0.10   & 16849.78           &  0.09   \\
$i^3\Pi_g^--c^3\Pi_u$         & 0-0      & P7      & 16787.90            & -0.10   & 16787.97           & -0.03   \\
                              &          & P14     & 16369.80            & -0.08   & 16369.84           & -0.04   \\
                              &          & Q3      & 17071.29            & -0.09   & 17071.28           & -0.10   \\
                              &          & Q9      & 16898.78            &  0.13   & 16898.69           &  0.04   \\
                              &          & R9      & 17150.62            &  0.17   & 17150.53           &  0.08   \\
                              &          & R12     & 17109.51            & -0.09   & 17109.64           &  0.04   \\
                              & 1-0      & P3      & 18525.88            &  0.22   & 18525.72           &  0.06   \\
                              &          & Q3      & 18598.23            & -0.27   & 18598.59           &  0.09   \\
                              & 2-1      & R1      & 18451.87            &  0.23   & 18451.64           &  0.00   \\
                              &          & R2      & 18464.46            &  0.13   & 18464.34           &  0.01   \\
                              & 2-2      & P10     & 16345.66            & -0.04   & 16345.72           &  0.02   \\
                              &          & R4      & 16864.08            &  0.21   & 16863.92           &  0.05   \\
                              &          & R8      & 16857.89            & -0.59   & 16858.46           & -0.02   \\
                              & 2-3      & Q3      & 15207.72            &  0.05   & 15207.75           &  0.08   \\
                              & 3-3      & P3      & 16517.58            &  0.06   & 16517.57           &  0.05   \\
                              &          & Q2      & 16599.70            &  0.05   & 16599.70           &  0.05   \\
                              &          & R5      & 16693.01            & -0.16   & 16693.13           & -0.04   \\
$j^3\Delta_g^+-c^3\Pi_u$      & 0-0      & P13     & 17309.93            & -0.02   & 17309.93           & -0.02   \\
                              &          & Q12     & 17667.66            &  0.00   & 17667.68           &  0.02   \\
                              & 1-0      & Q2      & 19060.17            &  0.14   & 19060.15           &  0.12   \\
                              & 1-1      & P10     & 17192.24            &  0.02   & 17192.26           &  0.04   \\
                              &          & Q7      & 17439.67            &  0.11   & 17439.65           &  0.09   \\
                              &          & R8      & 17722.82            & -0.04   & 17722.82           & -0.04   \\
                              & 2-1      & Q5      & 18907.85            & -0.11   & 18907.90           & -0.06   \\
                              & 2-2      & P9      & 17081.06            & -0.10   & 17081.14           & -0.02   \\
\hline
\end{tabular}
\end{table}

\newpage

\begin{table}
\begin{tabular}{@{}lllrrrr}
\multicolumn{7}{l}{\bf Table \ref{NewLines}. \rm (Continued.)} \\
\hline
 Electronic transition        &  Band    &  Line   &  \cite{Crosswhite}  &  O-C    &  This work         &  O-C    \\
\hline
$j^3\Delta_g^--c^3\Pi_u$      & 0-0      & P14     & 17182.21            & -0.01   & 17182.26           &  0.04   \\
                              &          & Q13     & 17564.20            & -0.02   & 17564.24           &  0.02   \\
                              &          & R4      & 17654.95            & -0.13   & 17655.06           & -0.02   \\
                              &          & R12     & 17922.15            &  0.22   & 17921.87           & -0.06   \\
                              & 1-0      & P7      & 18863.60            &  0.26   & 18863.37           &  0.03   \\	
                              &          & Q4      & 19066.14            & -0.18   & 19066.26           & -0.06   \\
                              & 1-1      & Q8      & 17423.59            &  0.10   & 17423.54           &  0.05   \\
                              &          & Q10     & 17431.73            & -0.09   & 17431.78           & -0.04   \\
                              &          & R9      & 17709.08            & -0.46   & 17709.58           &  0.04   \\
                              & 2-1      & P4      & 18786.75            & -0.14   & 18786.94           &  0.05   \\
                              &          & Q4      & 18903.82            & -0.28   & 18904.14           &  0.04   \\
                              &          & Q5      & 18903.31            & -0.22   & 18903.52           & -0.01   \\
                              & 2-2      & P3      & 17200.68            & -0.10   & 17200.72           & -0.06   \\
                              & 2-3      & R1      & 15783.87            & -0.25   & 15784.09           & -0.03   \\
                              & 3-3      & P3      & 17107.09            & -0.04   & 17107.11           & -0.02   \\
                              & 3-4      & Q2      & 15692.04            &  0.16   & 15691.96           &  0.08   \\
\hline
\end{tabular}
\end{table}


\clearpage

\begin{table}															
\caption{\label{NewLevels}
Optimal values $E_{nvN}$ of rovibronic energy levels (in cm$^{-1}$)
for various triplet electronic states of $D_2$ molecule obtained in 
the present work. The uncertainties of the $E_{nvN}$ value determination 
(one SD) are shown in brackets in units of last significant digit. 
$n_{\nu}$ --- the number of various spectral lines 
directly connected with certain level. $\Delta E$ is the difference
between energy level values obtained in the present work and those
reported in \cite{Crosswhite}.}

\begin{tabular}{@{}rr@{}c@{}lrrr@{}c@{}lrrr@{}c@{}lrr}
\hline                                                                                                                             
\multicolumn{16}{c}{$(1s\sigma 2s\sigma)a^3\Sigma_g^+$} \\
\hline
 & \multicolumn{5}{c}{v = 0} & \multicolumn{5}{c}{v = 1} & \multicolumn{5}{c}{v = 2} \\
\hline
 N & \multicolumn{3}{c}{ $E_{nvN}$ }& $n_{\nu}$ & $\Delta E$ &
     \multicolumn{3}{c}{ $E_{nvN}$ }& $n_{\nu}$ & $\Delta E$ &
     \multicolumn{3}{c}{ $E_{nvN}$ }& $n_{\nu}$ & $\Delta E$ \\
\hline
 0 &    0 & &              & 21  &  0     &  1814 &.& 908(5)  & 25    & -0.008 & 3560 &.& 726(6)  & 26    & -0.016   \\
 1 &   33 &.& 631(4)       & 43  & -0.011 &  1847 &.& 364(8)  & 55    & -0.014 & 3592 &.& 036(8)  & 56    & -0.016   \\
 2 &  100 &.& 770(6)       & 52  & -0.010 &  1912 &.& 172(5)  & 66    & -0.012 & 3654 &.& 573(5)  & 70    & -0.003   \\
 3 &  201 &.& 109(9)       & 47  & -0.009 &  2009 &.& 042(8)  & 63    & -0.002 & 3748 &.& 034(8)  & 64    & -0.004   \\
 4 &  334 &.& 333(6)       & 49  & -0.013 &  2137 &.& 628(6)  & 57    & -0.008 & 3872 &.& 103(6)  & 64    & -0.003   \\
\hline
\end{tabular}
\end{table}

\begin{table}

\begin{tabular}{@{}rr@{}c@{}lrrr@{}c@{}lrrr@{}c@{}lrr}
\multicolumn{16}{l}{\bf Table \ref{NewLevels}. \rm (Continued.)} \\
\hline                                                                                                                             
\multicolumn{16}{c}{$(1s\sigma 2s\sigma)a^3\Sigma_g^+$} \\
\hline
 & \multicolumn{5}{c}{v = 6} & \multicolumn{5}{c}{v = 7} & \multicolumn{5}{c}{v = 8} \\
\hline
 N & \multicolumn{3}{c}{ $E_{nvN}$ }& $n_{\nu}$ & $\Delta E$ &
     \multicolumn{3}{c}{ $E_{nvN}$ }& $n_{\nu}$ & $\Delta E$ &
     \multicolumn{3}{c}{ $E_{nvN}$ }& $n_{\nu}$ & $\Delta E$  \\
\hline
 0 &  9880 &.& 806(11) & 10         & -0.016 & 11299 &.& 845(16) & 6       & -0.045 & 12655 &.& 467(18) & 5       & -0.007 \\
 1 &  9907 &.& 775(10) & 26         &  0.005 & 11325 &.& 784(11) & 18      & -0.004 & 12680 &.& 356(12) & 12      & -0.036 \\
 2 &  9961 &.& 617(7)  & 32         & -0.017 & 11377 &.& 500(8)  & 18      &  0.000 & 12729 &.& 972(9)  & 13      & -0.052 \\
 3 & 10042 &.& 093(10) & 28         & -0.013 & 11454 &.& 821(12) & 14      & -0.010 & 12804 &.& 129(13) & 14      & -0.059 \\
\hline
\end{tabular}
\end{table}

\begin{table}
                                                                                                                                
\begin{tabular}{@{}rr@{}c@{}lrrr@{}c@{}lrrr@{}c@{}lrr}
\multicolumn{16}{l}{\bf Table \ref{NewLevels}. \rm (Continued.)} \\
\hline                                                                                                                             
\multicolumn{16}{c}{$(1s\sigma 2p\pi)c^3\Pi_u^-$} \\ 
\hline
 & \multicolumn{5}{c}{v = 3} & \multicolumn{5}{c}{v = 4} & \multicolumn{5}{c}{v = 5} \\ 
\hline
 N & \multicolumn{3}{c}{ $E_{nvN}$ }& $n_{\nu}$ & $\Delta E$ &
     \multicolumn{3}{c}{ $E_{nvN}$ }& $n_{\nu}$ & $\Delta E$ &
     \multicolumn{3}{c}{ $E_{nvN}$ }& $n_{\nu}$ & $\Delta E$  \\ 
\hline
 1 & 4727 &.& 022(9)  & 7             &  0.108 &  6226 &.& 366(19) & 3              &  0.144 & 7667 &.& 225(13) & 3            &  0.125 \\
 2 & 4782 &.& 112(9)  & 10            &  0.038 &  6279 &.& 454(11) & 7              &  0.096 & 7718 &.& 429(10) & 2            &  0.141 \\
 3 & 4864 &.& 307(13) & 11            &  0.073 &  6358 &.& 796(14) & 4              &  0.084 & 7794 &.& 944(13) & 2            &  0.116 \\
 4 & 4973 &.& 456(10) & 12            &  0.054 &  6464 &.& 117(13) & 4              &  0.063 & 7896 &.& 475(10) & 2            & -0.035 \\
\hline
\end{tabular}
\end{table}

\begin{table}

\begin{tabular}{@{}rr@{}c@{}lrrr@{}c@{}lrrr@{}c@{}lrr}
\multicolumn{16}{l}{\bf Table \ref{NewLevels}. \rm (Continued.)} \\
\hline                                                                                                                             
\multicolumn{16}{c}{$(1s\sigma 2p\pi)c^3\Pi_u^+$} \\ 
\hline
 & \multicolumn{5}{c}{v = 3} & \multicolumn{5}{c}{v = 4} & \multicolumn{5}{c}{v = 5} \\ \hline
 N & \multicolumn{3}{c}{ $E_{nvN}$ }& $n_{\nu}$ & $\Delta E$ &
     \multicolumn{3}{c}{ $E_{nvN}$ }& $n_{\nu}$ & $\Delta E$ &
     \multicolumn{3}{c}{ $E_{nvN}$ }& $n_{\nu}$ & $\Delta E$  \\ 
\hline
 1 & 4727 &.& 078(7)  & 14             &  0.052 & 6226 &.& 429(10) & 7              &  0.081 & 7667 &.& 297(13) & 2             &  0.053 \\
 2 & 4782 &.& 081(12) & 16             &  0.109 & 6279 &.& 553(10) & 9              &  0.007 & 7718 &.& 544(13) & 2             &  0.016 \\
 3 & 4864 &.& 446(8)  & 15             &  0.014 & 6359 &.& 007(11) & 7              &  0.133 & 7795 &.& 177(10) & 1             &  -     \\
 4 & 4973 &.& 606(14) & 9              &  0.084 & 6464 &.& 401(14) & 8              &  0.079 & 7896 &.& 833(13) & 1             &  -     \\
\hline
\end{tabular}
\end{table}

\begin{table}

\begin{tabular}{@{}rr@{}c@{}lrrr@{}c@{}lrrr@{}c@{}lrr}
\multicolumn{16}{l}{\bf Table \ref{NewLevels}. \rm (Continued.)} \\
\hline                                                                                                                             

\multicolumn{16}{c}{$(1s\sigma 3p\sigma)e^3\Sigma_u^+$} \\ 
\hline
 & \multicolumn{5}{c}{v = 0} & \multicolumn{5}{c}{v = 1} & \multicolumn{5}{c}{v = 2} \\ 
\hline
 N & \multicolumn{3}{c}{ $E_{nvN}$ }& $n_{\nu}$ & $\Delta E$ &
     \multicolumn{3}{c}{ $E_{nvN}$ }& $n_{\nu}$ & $\Delta E$ &
     \multicolumn{3}{c}{ $E_{nvN}$ }& $n_{\nu}$ & $\Delta E$  \\ 
\hline
 0 & 11649 &.& 190(14) & 5            & -0.010 & 13137 &.& 502(10) & 6       & -0.012 & 14558 &.& 258(12) & 6    &  0.002  \\
 1 & 11676 &.& 174(8)  & 12           & -0.004 & 13163 &.& 431(6)  & 12      & -0.011 & 14583 &.& 137(7)  & 14   &  0.003  \\
 2 & 11730 &.& 037(13) & 12           & -0.007 & 13215 &.& 170(9)  & 15      & -0.010 & 14632 &.& 770(11) & 13   &  0.000  \\
 3 & 11810 &.& 600(9)  & 17           & -0.020 & 13292 &.& 570(7)  & 16      & -0.010 & 14707 &.& 003(8)  & 15   & -0.003  \\
\hline
\end{tabular}
\end{table}

\begin{table}

\begin{tabular}{@{}rr@{}c@{}lrrr@{}c@{}lrrr@{}c@{}lrr}
\multicolumn{16}{l}{\bf Table \ref{NewLevels}. \rm (Continued.)} \\
\hline                                                                                                                             

\multicolumn{16}{c}{$(1s\sigma 3p\sigma)e^3\Sigma_u^+$} \\ 
\hline
 & \multicolumn{5}{c}{v = 6} & \multicolumn{5}{c}{v = 7} & \multicolumn{5}{c}{v = 8} \\ 
\hline
 N & \multicolumn{3}{c}{ $E_{nvN}$ }& $n_{\nu}$ & $\Delta E$ &
     \multicolumn{3}{c}{ $E_{nvN}$ }& $n_{\nu}$ & $\Delta E$ &
     \multicolumn{3}{c}{ $E_{nvN}$ }& $n_{\nu}$ & $\Delta E$  \\ 
\hline
 0 & 19546 &.& 641(11) & 7        & -0.001 & 20605 &.& 338(10) & 8         & -0.008 & 21575 &.& 310(13) & 9      & -0.030  \\
 1 & 19567 &.& 047(7)  & 16       & -0.007 & 20624 &.& 515(6)  & 15        & -0.015 & 21593 &.& 153(7)  & 19     & -0.013  \\
 2 & 19607 &.& 728(10) & 16       & -0.008 & 20662 &.& 787(9)  & 16        & -0.017 & 21628 &.& 745(10) & 19     & -0.005  \\
 3 & 19668 &.& 569(7)  & 16       &  0.001 & 20720 &.& 289(6)  & 14        & -0.009 & 21681 &.& 957(7)  & 20     & -0.007  \\
\hline
\end{tabular}
\end{table}

\begin{table}
                                                                                                                                
\begin{tabular}{@{}rr@{}c@{}lrrr@{}c@{}lrrr@{}c@{}lrr}
\multicolumn{16}{l}{\bf Table \ref{NewLevels}. \rm (Continued.)} \\
\hline                                                                                                                             

\multicolumn{16}{c}{$(1s\sigma 3p\pi)d^3\Pi_u^-$} \\ 
\hline
 & \multicolumn{5}{c}{v = 3} & \multicolumn{5}{c}{v = 4} & \multicolumn{5}{c}{v = 5} \\ \hline
 N & \multicolumn{3}{c}{ $E_{nvN}$ }& $n_{\nu}$ & $\Delta E$ &
     \multicolumn{3}{c}{ $E_{nvN}$ }& $n_{\nu}$ & $\Delta E$ &
     \multicolumn{3}{c}{ $E_{nvN}$ }& $n_{\nu}$ & $\Delta E$  \\ 
\hline
 1 & 21343 &.& 645(12) & 9       &  0.005 & 22769 &.& 442(12) & 8        & -0.012 & 24135 &.& 655(10) & 10    &  0.005  \\
 2 & 21396 &.& 902(9)  & 8       & -0.002 & 22820 &.& 678(8)  & 8        &  0.002 & 24184 &.& 948(7)  & 10    & -0.008  \\
 3 & 21476 &.& 477(12) & 8       & -0.007 & 22897 &.& 296(12) & 7        & -0.006 & 24258 &.& 610(10) & 9     &  0.000  \\
 4 & 21582 &.& 138(9)  & 8       & -0.008 & 22998 &.& 963(8)  & 8        & -0.013 & 24356 &.& 367(7)  & 9     & -0.017  \\
\hline
\end{tabular}
\end{table}

\begin{table}
                                                                                                                                
\begin{tabular}{@{}rr@{}c@{}lrrr@{}c@{}lrrr@{}c@{}lrr}
\multicolumn{16}{l}{\bf Table \ref{NewLevels}. \rm (Continued.)} \\
\hline                                                                                                                             
\multicolumn{16}{c}{$(1s\sigma 3p\pi)d^3\Pi_u^+$} \\ 
\hline
 & \multicolumn{5}{c}{v = 0} & \multicolumn{5}{c}{v = 1} & \multicolumn{5}{c}{v = 2} \\ 
\hline
 N & \multicolumn{3}{c}{ $E_{nvN}$ }& $n_{\nu}$ & $\Delta E$ &
     \multicolumn{3}{c}{ $E_{nvN}$ }& $n_{\nu}$ & $\Delta E$ &
     \multicolumn{3}{c}{ $E_{nvN}$ }& $n_{\nu}$ & $\Delta E$  \\ 
\hline
 1 & 16695 &.& 884(8)  & 8            & -0.014 & 18308 &.& 374(7)  & 12         &  0.006 & 19857 &.& 301(6)  & 15           & -0.011 \\
 2 & 16755 &.& 611(11) & 8            & -0.011 & 18365 &.& 854(9)  & 14         & -0.004 & 19912 &.& 998(10) & 12           &  0.002 \\
 3 & 16844 &.& 912(9)  & 8            & -0.002 & 18451 &.& 709(7)  & 13         & -0.019 & 19996 &.& 263(6)  & 16           & -0.013 \\
 4 & 16963 &.& 424(13) & 5            & -0.004 & 18565 &.& 166(10) & 13         &  0.004 & 20106 &.& 670(9)  & 14           &  0.000 \\
\hline
\end{tabular}
\end{table}

\begin{table}

\begin{tabular}{@{}rr@{}c@{}lrrr@{}c@{}lrrr@{}c@{}lrr}
\multicolumn{16}{l}{\bf Table \ref{NewLevels}. \rm (Continued.)} \\
\hline                                                                                                                             

\multicolumn{16}{c}{$(1s\sigma 3d\sigma)g^3\Sigma_g^+$} \\ 
\hline
 & \multicolumn{5}{c}{v = 3} & \multicolumn{5}{c}{v = 4} \\ 
\hline
 N & \multicolumn{3}{c}{ $E_{nvN}$ }& $n_{\nu}$ & $\Delta E$ &
     \multicolumn{3}{c}{ $E_{nvN}$ }& $n_{\nu}$ & $\Delta E$  \\ 
\hline
 0 & 20982 &.& 138(17) & 1       &  0.052 &       &-&         &        &         \\
 1 & 20996 &.& 599(14) & 3       &  0.091 &       &-&         &        &         \\
 2 & 21027 &.& 241(9)  & 6       &  0.029 & 22276 &.& 495(14) & 3      &  0.085  \\
 3 & 21075 &.& 984(15) & 6       &  0.156 & 22327 &.& 149(19) & 2      &  0.001  \\
\hline
\end{tabular}
\end{table}

\begin{table}

\begin{tabular}{@{}rr@{}c@{}lrrr@{}c@{}lrrr@{}c@{}lrr}
\multicolumn{16}{l}{\bf Table \ref{NewLevels}. \rm (Continued.)} \\
\hline
\multicolumn{16}{c}{$(1s\sigma 3d\pi)i^3\Pi_g^+$} \\ 
\hline
 & \multicolumn{5}{c}{v = 0} & \multicolumn{5}{c}{v = 1} & \multicolumn{5}{c}{v = 2} \\ 
\hline
 N & \multicolumn{3}{c}{ $E_{nvN}$ }& $n_{\nu}$ & $\Delta E$ &
     \multicolumn{3}{c}{ $E_{nvN}$ }& $n_{\nu}$ & $\Delta E$ &
     \multicolumn{3}{c}{ $E_{nvN}$ }& $n_{\nu}$ & $\Delta E$  \\ 
\hline
 1 & 16998 &.& 00(2)   & 4       &  0.024 & 18539 &.& 50(2)   & 5            &  0.040 &       &-&       &        &         \\
 2 & 17086 &.& 474(17) & 5       & -0.014 & 18617 &.& 862(19) & 6            &  0.008 & 20033 &.& 06(2) & 5      &  0.070  \\
 3 & 17202 &.& 627(19) & 5       & -0.007 & 18725 &.& 790(19) & 7            &  0.010 & 20130 &.& 70(3) & 3      &  0.080  \\
 4 & 17343 &.& 627(15) & 6       & -0.047 & 18860 &.& 377(15) & 7            &  0.033 & 20254 &.& 70(3) & 5      &  0.040  \\
\hline
\end{tabular}
\end{table}

\begin{table}

\begin{tabular}{@{}rr@{}c@{}lrrr@{}c@{}lrrr@{}c@{}lrr}
\multicolumn{16}{l}{\bf Table \ref{NewLevels}. \rm (Continued.)} \\
\hline                                                                                                                             

\multicolumn{16}{c}{$(1s\sigma 3d\delta)j^3\Delta_g^-$} \\ 
\hline
 & \multicolumn{5}{c}{v = 3} & \multicolumn{5}{c}{v = 4} & \multicolumn{5}{c}{v = 5} \\ 
\hline
 N & \multicolumn{3}{c}{ $E_{nvN}$ }& $n_{\nu}$ & $\Delta E$ &
     \multicolumn{3}{c}{ $E_{nvN}$ }& $n_{\nu}$ & $\Delta E$ &
     \multicolumn{3}{c}{ $E_{nvN}$ }& $n_{\nu}$ & $\Delta E$  \\ 
\hline
 2 & 21971 &.& 433(16) & 4       &  0.087 & 23370 &.& 57(3) & 3        &  0.070 & 24709 &.& 17(5)     & 2 &  0.070 \\
 3 & 22054 &.& 903(15) & 3       &  0.087 & 23448 &.& 57(3) & 3        &  0.090 &       &-&           &   &        \\
 4 & 22165 &.& 210(16) & 4       &  0.080 & 23551 &.& 78(4) & 2        &  0.060 &       &-&           &   &        \\
 5 & 22301 &.& 619(15) & 2       &  0.041 & 23679 &.& 34(4) & 2        &  0.030 &       &-&           &   &        \\
\hline
\end{tabular}
\end{table}

\begin{table}

\begin{tabular}{@{}rr@{}c@{}lrrr@{}c@{}lrrr@{}c@{}lrr}
\multicolumn{16}{l}{\bf Table \ref{NewLevels}. \rm (Continued.)} \\
\hline                                                                                                                             
\multicolumn{16}{c}{$(1s\sigma 4p\sigma)f^3\Sigma_u^+$} \\ 
\hline
 & \multicolumn{5}{c}{v = 0} & \multicolumn{5}{c}{v = 1} & \multicolumn{5}{c}{v = 2} \\ 
\hline
 N & \multicolumn{3}{c}{ $E_{nvN}$ }& $n_{\nu}$ & $\Delta E$ &
     \multicolumn{3}{c}{ $E_{nvN}$ }& $n_{\nu}$ & $\Delta E$ &
     \multicolumn{3}{c}{ $E_{nvN}$ }& $n_{\nu}$ & $\Delta E$  \\ 
\hline
 0 & 20546 &.& 04(3) & 2       &  -0.010 & 22098 &.& 279(15) & 3        & -0.019 & 23584 &.& 958(15) & 3     & -0.008 \\
 1 & 20574 &.& 22(2) & 3       &   0.000 & 22125 &.& 052(8)  & 6        & -0.002 & 23610 &.& 648(8)  & 8     & -0.008 \\
 2 & 20630 &.& 49(2) & 3       &  -0.020 & 22178 &.& 513(12) & 6        &  0.007 & 23661 &.& 968(13) & 8     & -0.012 \\
 3 & 20714 &.& 33(2) & 4       &  -0.010 & 22258 &.& 589(10) & 6        &  0.011 & 23738 &.& 768(9)  & 8     & -0.008 \\
\hline
\end{tabular}
\end{table}

\begin{table}

\begin{tabular}{@{}rr@{}c@{}lrrr@{}c@{}lrrr@{}c@{}lrr}
\multicolumn{16}{l}{\bf Table \ref{NewLevels}. \rm (Continued.)} \\
\hline                                                                                                                             

\multicolumn{16}{c}{$(1s\sigma 4p\pi)k^3\Pi_u^+$} \\ 
\hline
 & \multicolumn{5}{c}{v = 0} & \multicolumn{5}{c}{v = 1} & \multicolumn{5}{c}{v = 2} \\ 
\hline
 N & \multicolumn{3}{c}{ $E_{nvN}$ }& $n_{\nu}$ & $\Delta E$ &
     \multicolumn{3}{c}{ $E_{nvN}$ }& $n_{\nu}$ & $\Delta E$ &
     \multicolumn{3}{c}{ $E_{nvN}$ }& $n_{\nu}$ & $\Delta E$  \\ 
\hline
 1 & 22353 &.& 203(13) & 4       & -0.003 & 23944 &.& 921(16) & 5           & -0.011 & 25472 &.& 366(8)  & 5   & -0.026 \\
 2 & 22413 &.& 471(18) & 3       & -0.011 & 24003 &.& 076(18) & 6           &  0.004 & 25528 &.& 246(10) & 6   &  0.004 \\
 3 & 22503 &.& 474(14) & 4       & -0.004 & 24089 &.& 428(16) & 6           &  0.012 & 25611 &.& 819(8)  & 5   &  0.041 \\
 4 & 22622 &.& 76(2)   & 3       & -0.019 & 24205 &.& 036(19) & 6           & -0.006 & 25723 &.& 05(8)   & 2   & -0.030 \\
\hline
\end{tabular}
\end{table}

\begin{table}

\begin{tabular}{@{}rr@{}c@{}lrrr@{}c@{}lrrr@{}c@{}lrr}
\multicolumn{16}{l}{\bf Table \ref{NewLevels}. \rm (Continued.)} \\
\hline                                                                                                                             
 & \multicolumn{5}{c}{$(1s\sigma 4d\pi)r^3\Pi_g^-$}       &
   \multicolumn{5}{c}{$(1s\sigma 4d\pi)r^3\Pi_g^+$}       & 
   \multicolumn{5}{c}{$(1s\sigma 4d\delta)s^3\Delta_g^-$} \\
\hline
 & \multicolumn{5}{c}{v = 0} &
   \multicolumn{5}{c}{v = 0} & 
   \multicolumn{5}{c}{v = 0} \\ 
\hline
 N & \multicolumn{3}{c}{ $E_{nvN}$ }& $n_{\nu}$ & $\Delta E$ &
     \multicolumn{3}{c}{ $E_{nvN}$ }& $n_{\nu}$ & $\Delta E$ &
     \multicolumn{3}{c}{ $E_{nvN}$ }& $n_{\nu}$ & $\Delta E$  \\ 
\hline
 1 & 22489 &.& 55(4) & 3 &  0.000 &       &-&       &   &        &       &-&       &     &        \\
 2 & 22517 &.& 93(3) & 3 & -0.000 & 22603 &.& 44(3) & 4 & -0.030 & 22673 &.& 64(3) & 3   &  0.040 \\
 3 & 22576 &.& 61(4) & 3 & -0.020 & 22693 &.& 53(3) & 4 &  0.000 & 22788 &.& 44(2) & 3   & -0.030 \\
 4 & 22664 &.& 80(3) & 3 & -0.020 & 22810 &.& 90(3) & 3 & -0.060 & 22931 &.& 97(3) & 3   & -0.030 \\
\hline
\end{tabular}
\end{table}

\begin{table}

\begin{tabular}{@{}rr@{}c@{}lrrr@{}c@{}lrrr@{}c@{}lrr}
\multicolumn{16}{l}{\bf Table \ref{NewLevels}. \rm (Continued.)} \\
\hline
\multicolumn{5}{c}{$(1s\sigma 5d\sigma)q^3\Sigma_g^+$} \\ 
\hline
 & \multicolumn{5}{c}{v = 0} \\ 
\hline
 N & \multicolumn{3}{c}{ $E_{nvN}$ }& $n_{\nu}$ & $\Delta E$  \\ 
\hline
 2 & 24936 &.& 06(2) & 2       & -0.010 \\
 3 & 24963 &.& 90(2) & 2       & -0.040 \\
 4 & 25021 &.& 59(3) & 1       & -0.030 \\
 5 & 25108 &.& 37(3) & 1       & -0.030 \\
\hline
\end{tabular}
\end{table}

\begin{table}

\begin{tabular}{@{}rr@{}c@{}lrrr@{}c@{}lrrr@{}c@{}lrr}
\multicolumn{16}{l}{\bf Table \ref{NewLevels}. \rm (Continued.)} \\
\hline                                                                                                                             

\multicolumn{11}{c}{$(1s\sigma 7p\pi)^3\Pi_u^-$} & \multicolumn{5}{c}{$(1s\sigma 7d\pi)^3\Pi_u^-$}\\ 
\hline
 & \multicolumn{5}{c}{v = 0} & \multicolumn{5}{c}{v = 1} &  \multicolumn{5}{c}{v = 0} \\ 
\hline
 N & \multicolumn{3}{c}{ $E_{nvN}$ }& $n_{\nu}$ & $\Delta E$ &
     \multicolumn{3}{c}{ $E_{nvN}$ }& $n_{\nu}$ & $\Delta E$ &
     \multicolumn{3}{c}{ $E_{nvN}$ }& $n_{\nu}$ & $\Delta E$  \\ 
\hline
 1 & 27149 &.& 29(5) & 1       &  -0.010 & 28727 &.& 59(5) & 1    & -0.010 &       &-&       &           &        \\
 2 & 27207 &.& 37(5) & 1       &  -0.010 & 28784 &.& 23(5) & 1    & -0.010 & 27136 &.& 91(5) & 1         &  0.000 \\
 3 &       &-&       &         &         & 28868 &.& 69(5) & 1    &  0.000 & 27163 &.& 62(5) & 1         &  0.010 \\
 4 & 27411 &.& 74(5) & 1       &  -0.010 & 28980 &.& 94(5) & 1    & -0.010 & 27221 &.& 01(5) & 1         & -0.030 \\
\hline
\end{tabular}
\end{table}
}

\end{document}